\documentclass[lettersize,journal]{IEEEtran}
\usepackage{amsmath,amsfonts}
\usepackage{algorithmic}
\usepackage{algorithm}
\usepackage{array}
\usepackage{textcomp}
\usepackage{stfloats}
\usepackage{url}
\usepackage{verbatim}
\usepackage{graphicx}
\usepackage{cite}
\usepackage{subfigure,dcolumn}
\usepackage[T2A,T1]{fontenc}
\usepackage[russian,english]{babel}
\usepackage{multirow}
\usepackage{amssymb}
\usepackage{booktabs}
\usepackage{color}
\usepackage{xspace}
\usepackage{bm}

\usepackage[colorlinks=true,
    linkcolor=blue,
    urlcolor=blue,citecolor=blue]{hyperref}

\usepackage[T1]{fontenc}
\usepackage[misc]{ifsym}

\usepackage{listings}
\newcommand{\ie}{\textit{i}.\textit{e}.\xspace}
\newcommand{\eg}{\textit{e}.\textit{g}.\xspace}

\newcommand{\std}[1]{\tiny #1}
\newcommand{\mat}[1]{\boldsymbol{#1}}

\newcommand{\modelname}{LangMamba\xspace}
\newcommand{\generator}{SEED\xspace}
\newcommand{\lossname}{LangDA\xspace}
\newcommand{\aename}{LangAE\xspace}

\newlength\savewidth\newcommand\shline{\noalign{\global\savewidth\arrayrulewidth
  \global\arrayrulewidth 1.5pt}\hline\noalign{\global\arrayrulewidth\savewidth}}

\begin{document}

\title{LangMamba: A Language-driven Mamba Framework for Low-dose CT Denoising with Vision-language Models}

\author{Zhihao~Chen,~Tao~Chen,~Chenhui~Wang,~Qi~Gao,~Huidong~Xie,~Chuang~Niu,~\IEEEmembership{Member,~IEEE},\\~Ge~Wang,~\IEEEmembership{Life Fellow,~IEEE},~and~Hongming~Shan,~\IEEEmembership{Senior~Member,~IEEE}

\thanks{This work was supported in part by the National Natural Science Foundation of China (Nos. 62471148 and 62101136) and the National Institutes of Health (Nos. R01CA237267, R01HL151561, R01EB031102, and R01EB032716).}
\thanks{This work involved human subjects or animals in its research.   The authors confirm that all human/animal subject research procedures and protocols are exempt from review board approval.}
\thanks{Z. Chen, T. Chen,~C. Wang and Q. Gao are with the Institute of Science and Technology for Brain-inspired Intelligence, Fudan University, Shanghai 200433, China (e-mail: zhihaochen21@m.fudan.edu.cn; 21110850023@m.fudan.edu.cn; chenhuiwang21@m.fudan.edu.cn; qgao21@m.fudan.edu.cn)}
\thanks{H. Xie is with the Department of Biomedical Engineering, Yale University, New Haven, CT, 06510, USA (e-mail:huidong.xie@yale.edu)}
\thanks{C. Niu and G. Wang are with Biomedical Imaging Center, Center for Biotechnology and Interdisciplinary Studies, Department of Biomedical Engineering,
Rensselaer Polytechnic Institute, Troy, NY 12180, USA (e-mail: niuc@rpi.edu; wangg6@rpi.edu)}
\thanks{H. Shan is with the Institute of Science and Technology for Brain-inspired Intelligence, MOE Frontiers Center for Brain Science and Key Laboratory of Computational Neuroscience and Brain-Inspired Intelligence (Ministry of Education), Fudan University, Shanghai 200433, China, and also with the Shanghai Center for Brain Science and Brain-inspired Technology, Shanghai 201210, China (e-mail: hmshan@fudan.edu.cn).}
}

\markboth{IEEE Transactions on Radiation and Plasma Medical Sciences,~Vol.~xx, No.~x,~2025}%
{Shell \MakeLowercase{\textit{et al.}}: A Sample Article Using IEEEtran.cls for IEEE Journals}

\maketitle

\begin{abstract}
Low-dose computed tomography (LDCT) reduces radiation exposure but often degrades image quality, potentially compromising diagnostic accuracy. 
Existing deep learning-based denoising methods focus primarily on pixel-level mappings, overlooking the potential benefits of high-level semantic guidance.
Recent advances in vision-language models (VLMs) suggest that language can serve as a powerful tool for capturing structured semantic information, offering new opportunities to improve LDCT reconstruction.
In this paper, we introduce \modelname, a Language-driven Mamba framework for LDCT denoising that leverages VLM-derived representations to enhance supervision from normal-dose CT (NDCT).
\modelname follows a two-stage learning strategy. 
First, we pre-train a Language-guided AutoEncoder (\aename) that leverages frozen VLMs to map NDCT images into a semantic space enriched with anatomical information. 
Second, we synergize \aename with two key components to guide LDCT denoising: Semantic-Enhanced Efficient Denoiser (\generator), which enhances NDCT-relevant local semantic while capturing global features with efficient Mamba mechanism, and Language-engaged Dual-space Alignment (\lossname) Loss, which ensures that denoised images align with NDCT in both perceptual and semantic spaces.
Extensive experiments on two public datasets demonstrate that \modelname outperforms conventional state-of-the-art methods, significantly improving detail preservation and visual fidelity. 
Remarkably, \aename exhibits strong generalizability to unseen datasets, thereby reducing training costs.
Furthermore, \lossname loss improves explainability by integrating language-guided insights into image reconstruction and offers a plug-and-play fashion. 
Our findings shed new light on the potential of language as a supervisory signal to advance LDCT denoising. 
The code is publicly available on \url{https://github.com/hao1635/LangMamba}.
\end{abstract}

\begin{IEEEkeywords}
CT denoising, Vision-language model, Mamba, State space models, Vector quantization, Explainability.
\end{IEEEkeywords}

\section{Introduction}
Computed tomography (CT) is a widely used medical imaging technique that uses a rotating X-ray tube and detectors to measure tissue attenuation, producing detailed anatomical images~\cite{lei2023ct}.
However, the ionizing radiation associated with CT scans may pose health risks, including tissue damage, DNA mutation, and an increased cancer risk~\cite{chen2024deep}.
Low-dose CT~(LDCT) reduces the radiation exposure, making it widely used in applications that require repeated scans such as lung cancer screening~\cite{lewis2015low}. 
However, LDCT negatively affects image quality and may compromise diagnostic accuracy~\cite{hein2024ppfm}. 
Therefore, reconstructing high-quality images from noisy input without compromising diagnostic performance remains a significant challenge. 

While existing deep learning-based methods~\cite{shan20183, wang2025self, lu2024pridediff, litformer} have shown promise in LDCT denoising,
they mostly focus on pixel-level mappings, which overlook high-level semantics that could guide more accurate reconstruction~\cite{sun2023coser}.
This limitation results in inconsistent reconstruction with the NDCT image quality.
Recent advances in vision-language models (VLMs)~\cite{clip,blip} have demonstrated the capability of textual representations to capture semantic knowledge, enabling effective guidance of various vision tasks, such as image understanding~\cite{chen2023iqagpt,yu2024spae} and segmentation~\cite{llmseg}.
VLMs introduce a novel strategy to enhance LDCT denoising by leveraging language as a supervisory signal to capture NDCT-relevant semantic information, thereby complementing traditional pixel-wise supervision.
Despite this potential, the direct integration of large-scale VLMs into high-resolution medical imaging remains computationally prohibitive, limiting their clinical applicability.
Therefore, developing efficient strategies to utilize semantic information extracted from VLMs is a key challenge, with the goal of achieving VLM-supervision and optimizing denoising performance without introducing excessive computational burdens.

To address this challenge, we propose \textit{\modelname}, a novel \textbf{Lang}uage-driven \textbf{Mamba} framework for LDCT denoising.
\modelname ingeniously leverages VLM-derived representations to enhance the supervision provided by NDCT images, guiding the denoising process with both local and global semantic context. 
Our approach is built upon a two-stage learning strategy:
In the first stage, we pre-train a Language-guided AutoEncoder (\aename), which is built upon a vector quantized generative adversarial network (VQGAN)~\cite{vq-gan}. 
Unlike the original VQGAN, we replace the learnable codebook with pretrained token embeddings from a frozen VLM~\cite{LQAE} and employ a pyramid semantic loss~\cite{yu2024spae} to align quantized tokens with anatomical information across multiple scales, thereby enhancing the semantic representation.
In the second stage, we leverage the \aename to develop a denoising model incorporating two novel components: a Semantic-Enhanced Efficient Denoiser (\generator) and a Language-engaged Dual-space Alignment (\lossname) loss.
\generator transfers frozen \aename encoder to enhance NDCT-relevant local context and employs decoder with efficient Mamba attention~(EMA) blocks for global information extraction with linear complexity.
\lossname loss, also guided by \aename, provides additional supervision by minimizing the discrepancy between denoised CT and NDCT images in terms of both the continuous features and discrete semantic token embeddings.
Crucially, the VLM is not directly involved in the training or inference of denoising, significantly reducing computational and deployment costs.

Through its resource-efficient integration of VLM-derived capabilities, \modelname effectively suppresses noise, preserves finer details, and enhances visual fidelity that closely resembles NDCT, all within a single framework.
Importantly, the first-stage \aename requires no retraining on new datasets, demonstrating generalization to unseen data and significantly reducing deployment time in new scenarios.
Moreover, leveraging the extensive capabilities of LLMs, our \modelname helps understand the anatomical semantic information in the denoised image with quantized text tokens during the denoising process, which can improve explainability in clinical applications and make it more trustworthy for radiologists.

Our contributions are summarized as follows:
\begin{itemize}
\item We propose \modelname, a generalizable LDCT denoising framework consisting of a two-stage learning strategy: (1) VLM-guided \aename pretraining to encode semantic information, and (2) a denoising model that incorporates \generator and \lossname loss. This design efficiently fuses semantic information derived from VLMs to enhance LDCT denoising.

\item \aename encodes CT images into continuous features and discretizes them into semantic-relevant text tokens via a frozen VLM codebook.

\item \generator enhances NDCT-relevant local semantic information through the \aename encoder while capturing global features with an efficient Mamba mechanism.

\item \lossname loss, also guided by the frozen \aename, aligns denoised CT and NDCT in both continuous perceptual and discrete semantic spaces.

\item  Our extensive experimental results on two public datasets demonstrate that \modelname efficiently preserves finer details, enhances visual fidelity, and provides language-level explainability.
To the best of our knowledge, this is the first work to apply VLM for LDCT denoising.
\end{itemize}

We note that a preliminary version of this work, \aename was published in the IEEE International Conference on Bioinformatics and Biomedicine 2024~\cite{chen2024leda}. 
In this paper, we further extend \aename~\cite{chen2024leda} with the following major improvements. 
First, we conduct a more in-depth analysis of \aename and integrate it into a novel \generator that efficiently captures global and local information by leveraging the Mamba mechanism. 
Second, building on this foundation, we propose a novel denoising framework, \modelname, which synergistically combines these advancements to further improve denoising performance.
Third, we extend our method to another commonly used public dataset and explore the generalizability of \aename, providing additional validation of the proposed model's effectiveness.

\begin{figure*}[!t]
\centering
\includegraphics[width=1\linewidth]{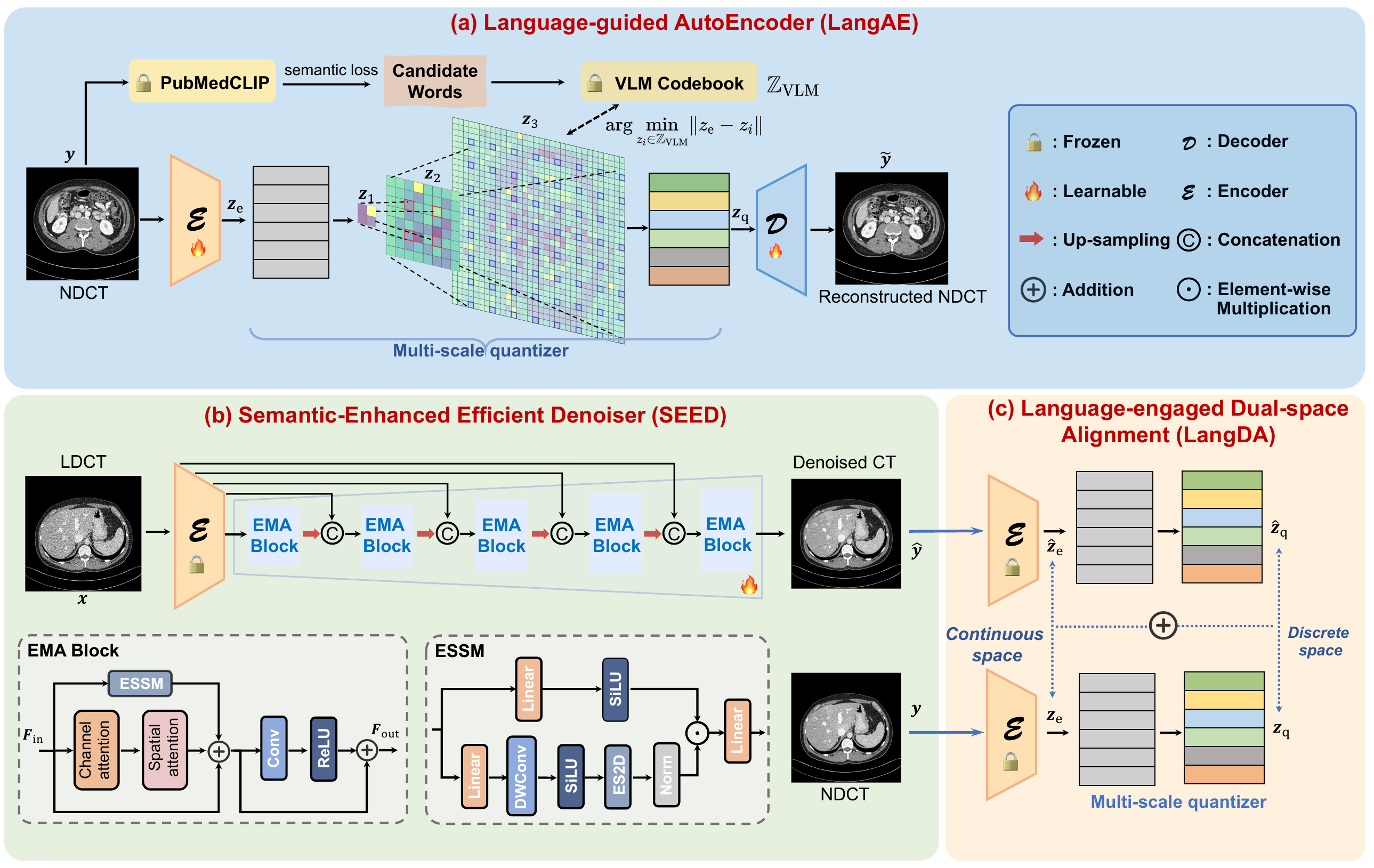}
\caption{Overall architecture of proposed language-driven Mamba framework~(\modelname); (a) Architecture of language-guided autoencoder~(\aename); (b) Architecture of semantic-enhanced efficient denoiser~(\generator); (c) Architecture of language-engaged dual-space alignment~(\lossname) loss.}
\label{fig:network}
\end{figure*}

The remainder of this paper is organized as follows. Sec.~\ref{sec:related_works} reviews the related literature. Sec.~\ref{sec:method} presents the proposed methodology. Sec.~\ref{sec:experiment} provides comprehensive
experimental results on the two public datasets, followed by a discussion in Sec.~\ref{sec:discussion}. 
Finally, Sec.~\ref{sec:conclusion} concludes this paper with a summary of key findings.

\section{Related works}
\label{sec:related_works}

\subsection{Vision Quantization}
VQ-VAE~\cite{van2017neural} encodes data into a discrete latent space by applying vector quantization with the use of a codebook. 
Building on this, VQGAN~\cite{vq-gan} improves reconstruction quality by incorporating adversarial and perceptual objectives. 
The resulting discrete latent representations, commonly referred to as tokens, are widely utilized in training generative models~\cite{SD}. 
In conventional VQGANs, the codebooks are learned jointly with the encoder and decoder, which limits their interpretability in natural language contexts. 
To address this, LQAE~\cite{LQAE} replaces the learned codebook with frozen word embeddings from BERT~\cite{kenton2019bert}, enabling alignment with an English vocabulary. 
However, LQAE tokens often fail to capture semantic concepts inherent in images and exhibit inferior reconstruction quality compared to models with learned codebooks. 
SPAE~\cite{yu2024spae} model maps input data to semantically meaningful tokens derived from a multilingual vocabulary while maintaining the high reconstruction quality of VQGAN. 
In our \aename, we use the language as additional supervision in the autoencoder to ensure that the quantized text tokens are semantically rich and aligned with anatomical information revealed by CT.

\subsection{State Space Models (SSMs)}
State Space Models (SSMs), rooted in linear time-invariant~(LTI) systems~\cite{gu2021efficiently, smith2022simplified}, have emerged as efficient deep learning alternatives for modeling long-range dependencies. 
LTI systems transform a sequence $x(t) \in \mathbb{R}$ to $y(t) \in \mathbb{R}$ via an implicit latent state $\mathbf{h}(t) \in \mathbb{R}^{N}$, governed by a linear ordinary differential equation (ODE):

\begin{equation}
\begin{aligned}
\label{eq:ssm}
    \mat{h^{\prime}}(t)&={\mat{A}}\mat{h(t)}+{\mat{B}}{x(t)},\\
    y(t)&={\mat{C}}\mat{h}(t)+{D}{x(t)},
\end{aligned}
\end{equation}
where $N$ is the state size, ${\mat{A}} \in \mathbb{R}^{N\times N}$, ${\mat{B}} \in \mathbb{R}^{N \times 1}$, ${\mat{C}} \in \mathbb{R}^{1\times N}$, and ${D} \in \mathbb{R}$.

Subsequently, a discretization process is applied to convert Eq.~\eqref{eq:ssm} into a form suitable for practical deep learning algorithms. 
Specifically, let $\rm \Delta$ represent the timescale parameter used to transform the continuous matrices ${{\mat{A}}}$, ${{\mat{B}}}$  into their discrete counterparts, $\overline{{\mat{A}}}$, $\overline{{\mat{B}}}$.  
The zero-order hold (ZOH) is used to approximate discretization, which is defined as:
\begin{equation}
\begin{aligned}
    \overline{\mat{A}} &= \mathrm{exp}({\mathrm{\Delta} \mat{A}}),\\
    \overline{\mat{B}}&=({ \mathrm{\Delta} \mat{A}})^{-1}({ \mathrm{exp}(\mat{A})}-\mat{I})\cdot {\mathrm{\Delta} \mat{B}}.
\end{aligned}
\end{equation}

Following the discretization, the resulting discrete form of Eq.~\eqref{eq:ssm} with step size $ \Delta$ can be expressed in the following recurrent neural network form:
\begin{equation}
\begin{aligned}
\label{eq:discret-ssm}
    \mat{h}_k&=\overline{\mat{A}}\mat{h}_{k-1}+\overline{\mat{B}}x_k,\\
    y_k&={\mat{C}}\mat{h}_k+{D}x_k.
\end{aligned}
\end{equation}

Additionally, Eq.~\eqref{eq:discret-ssm} can be mathematically transformed into the following CNN form, enabling the efficient training:
\begin{equation}
\begin{aligned}
\label{eq:cnn-form}
\overline{\mat{K}}&\triangleq(\mathrm{\textbf{C}} \overline{\mathrm{\textbf{B}}},{\rm{\textbf{C}}}\overline{\mat{A}}\overline{\mat{B}},\cdots,{\rm{\textbf{C}}}{\overline{\mat{A}}}^{L-1}\overline{\mat{B}}),\\
{\mat{y}}&={\mat{x}} \circledast \overline{\mat{K}},
\end{aligned}
\end{equation}
where $L$ is the length of the input sequence, $\circledast$ denotes convolution operation, and $\overline{\mat{K}} \in \mathbb{R}^L$ is a structured convolution kernel.

Structured state-space sequence models (S4)~\cite{gu2021efficiently}, use diagonalized parameter matrices for more efficient sequence modeling. 
The recent Mamba~\cite{gu2023mamba} extends S4 with a selection mechanism that dynamically adjusts $\overline{\mat{B}}$, ${\mat{C}}$, and $\rm \Delta$ based on inputs, filtering irrelevant information. 
To apply SSM to vision tasks, the 2D-Selective-Scan (SS2D) module~\cite{liu2024vmamba,guo2025mambair} reformulates SSM using convolutions, extending the kernel from 1D to 2D via outer product, enabling selective SSM (S6)~\cite{gu2023mamba} for vision without losing its advantages.
In this work, we integrate a lightweight Mamba mechanism~\cite{pei2024efficientvmamba} that leverages skipping sampling and combines processed patches for efficient global feature extraction.

\section{Methodology}
\label{sec:method}
\subsection{Overview of \modelname}
Fig.~\ref{fig:network} presents the overall framework of the proposed \modelname, which follows a two-stage learning strategy.
First, we pretrain the \aename using a frozen medical VLM~\cite{pubmedclip}, which encodes an NDCT image $\mat{y}$ into a continuous perceptual space and subsequently quantizes it into a multi-scale semantic space before decoding it back to a reconstructed NDCT image $\mat{\tilde{y}}$, as illustrated in Fig.~\ref{fig:network}(a).
Then, building on LangAE, we develop our denoising model by leveraging its encoder and multi-scale quantizer.
Given an LDCT image, $\mat{x} \in \mathbb{R}^{1 \times H \times W}$, where $H\times W$ denotes the image size, $\mat{x}$ is passed through the \generator.
\generator consists of the first five scales of dense feature maps extracted from the LangAE encoder and a decoder that incorporates EMA blocks, producing a denoised CT image $\mat{\hat{y}} \in \mathbb{R}^{1 \times H \times W}$.
To enhance supervision and explore language-level explainability, denoised CT $\mat{\hat{y}}$ and NDCT $\mat{y}$ are passed to the pretrained \aename to compute a \lossname loss in both continuous perceptual and discrete semantic spaces. 

Next, we detail the training process of LangAE in Subsec.~\ref{sec:autoencoder}, followed by a comprehensive introduction of the two key components of our denoising model: \generator (Subsec.~\ref{sec:mcunet}) and \lossname loss (Subsec.~\ref{sec:leda}).

\subsection{Language-guided Autoencoder (\aename)}
\label{sec:autoencoder}
To effectively incorporate the semantic information extraction capabilities of VLMs into LDCT denoising, we pretrain a language-guided autoencoder using NDCT images.
The autoencoder is built upon a VQGAN~\cite{vq-gan} architecture 
that consists of an encoder $\mathcal{E}$, a decoder $\mathcal{D}$, and a codebook $\mathbb{Z}$. Given an NDCT image $\mat{y}$, $\mathcal{E}$ encodes it to produce high-level perceptual features $\mathcal{E}(\mat{y})$ = $\mat{z}_\mathrm{e} \in \mathbb{R}^{C \times H^{\prime} \times W^{\prime}}$ in the continuous space.
In the conventional vector quantizer, $\mat{z}_\mathrm{e}$ is then quantized into the discrete space through nearest neighbors lookup: $ \mat{z}_\mathrm{q} =\arg \min _{z_{i} \in \mathbb{Z}}\left\|\mat{z}_\mathrm{e}-\mat{z_{i}}\right\|$. However, these discrete representations lack concrete semantic information, which is vital for medical imaging tasks.

In order to address this semantic gap, we make two modifications to the original VQGAN quantizer using a pre-trained VLM, as shown in Fig.~\ref{fig:network}(a). 
First, we replace the learned codebook $\mathbb{Z}$  with a fixed codebook $\mathbb{Z_\mathrm{VLM}}=\{(t, \mat{e}(t)) \mid t \in \mathbb{V}\}$ from the pretrained medical VLM, PubMedCLIP~\cite{pubmedclip}, where $\mathbb{V}$ is the language vocabulary and $\mat{e}(\cdot)$ produces the text embedding for a token $t$. 
It enables the encoder to map perceptual features into a discrete space delineated by text token embeddings, making the quantized representation directly convertible into text. 
However, these text tokens are random and may not be interpretable by humans~\cite{LQAE}. 
Second, we use a pyramid semantic loss from SPAE~\cite{yu2024spae} in a pyramid quantizer to learn hierarchical semantic information. 
As shown in Fig.~\ref{fig:network}(a), the quantized token pyramid consists of 3 layers, with tokens at layer $l$ are denoted as $\mat{t}_{l} \in \mathbb{V}^{h_{l} \times w_{l}}$. We define a set $\mathbf{P}(l)$ that includes the selected positions $(x, y)$ for quantization at layer $l$ using the dilation sampling operation, which are marked in the quantizer of Fig.~\ref{fig:network}(a).

Concretely, we build candidate token pools for each layer $\mathbf{C}_{l}(\mat{y})=\{t \in \mathbb{V} \mid s(\mat{y}, t) \geq \rho_{l}\}$ according to the similarity score calculated by PubMedCLIP, where $ s(\mat{y}, t)$ is the similarity score and $\rho_{l}$ is a threshold.
The pyramid semantic loss encourages the semantic similarity between an input NDCT image and each token in $\mathbf{C}_{l}(\mat{y})$, defined as:
\begin{align}
    \mathcal{L}_{\text {sem}}= \underset{l \in [1, D]}{\mathbb{E}}\;\underset{\mathbf{z}_{l}}{\mathbb{E}}\; \underset{c \in \mathbf{C}_{l}(\mat{y})}{\mathbb{E}}-\log \frac{\exp(\|(\mat{z}_{l}-\mat{e}(c) \|_{2}^{2})}{\sum\limits_{k \in \mathbb{V}}\exp (-\|\mat{z}_{l}\!-\!\mat{e}(k)\|_{2}^{2})},
\end{align}
where $\mat{z}_{l}$ is the $l$-th layer embedding, calculated from
\begin{align}
   \mat{z}_{l}=\boldsymbol{z}+\sum_{i=1}^{l-1} \mathbf{1}_{(x, y) \in \mathbf{P}(i)}\Big(\boldsymbol{z}-\boldsymbol{e}\left(t_{i}\right)\Big),
\end{align}
where $\mathbf{P}(l)$ is a set including the positions $(x, y)$ for quantization at layer $l$, as shown in Fig.~\ref{fig:network}(a). 
Each element $\mat{z} \in \mat{z}_\mathrm{e} $  is passed through the quantizer, which assigns it to the closest entry in a codebook through $t_{l}=\arg \min _{t \in \mathbb{Z}_{\mathrm{VLM}}}\left\|\mathbf{z}_{l}-\mathbf{e}(t)\right\|_{2}^{2}$. 
Final quantized token embeddings are obtained from the first $l$ layers ($l=3$) and given by the average of the existing token embeddings as
\begin{align}
    \mat{z}_\mathrm{q}=\boldsymbol{z}_{\leq l}=\frac{\sum\limits_{i=1}^{l} \mathbf{1}_{(x, y) \in \mathbf{P}(i)} \boldsymbol{e}\left(t_{i}\right)}{\sum\limits_{i=1}^{l} \mathbf{1}_{(x, y) \in \mathbf{P}(i)}}.
\end{align}

The total training loss of the \aename encompasses both the VQGAN loss~\cite{vq-gan,LQAE} and the pyramid semantic loss: 
\begin{align}
&\mathcal{L}_{\text{total}}=\mathcal{L}_{\text{VQGAN}}+\alpha\omega\mathcal{L}_{\text {sem}},
\end{align}
where 
\begin{align}
\mathcal{L}_{\text{VQGAN}}&=\left\|\mat{y}-\tilde{\mat{y}}\right\|_{2}^{2}+\beta \sum_{l=1}^{D}\|\mat{z}-\operatorname{sg}(\mat{z}_\mathrm{q})\|_{2}^{2} \notag\\
&+\gamma \mathcal{L}_{\mathrm{GAN}}+\eta \mathcal{L}_{\text {Perceptual}}, 
\end{align}
where $\mathcal{L}_{\mathrm{GAN}}$ and $ \mathcal{L}_{\text {Perceptual}}$ 
are GAN~\cite{vq-gan} and perceptual~\cite{johnson2016perceptual} losses respectively. $\operatorname{sg}$ is the abbreviation of stop gradient. Following~\cite{yu2024spae}, the hyperparameters $\alpha$, $\beta$, $\lambda$, 
 and $\eta$ are set to $0.3$, $0.3$, $0.1$, and $0.1$, respectively. $\omega=(\mathcal{L}_{\text{VQGAN}}/{\mathcal{L}_{\text {sem}}})$ is the dynamic weight without gradient backpropagation.

\begin{figure}[t]
\centering
\includegraphics[width=1\linewidth]
{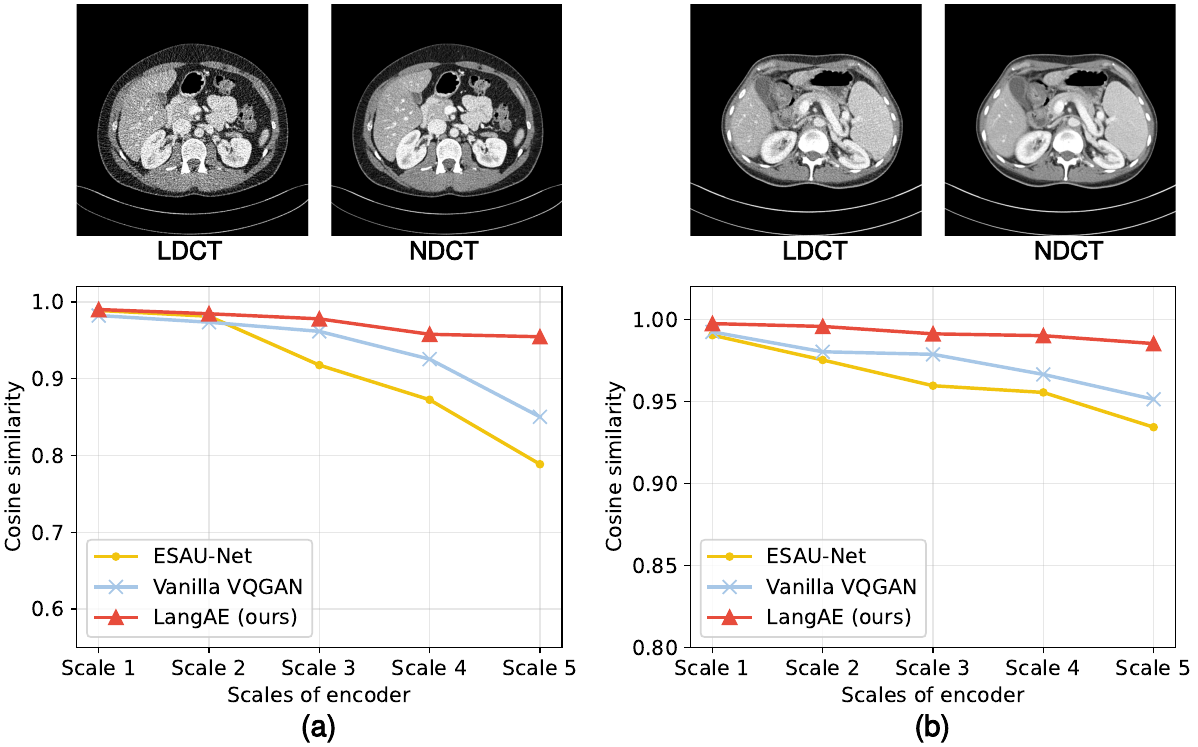}
\caption{Feature similarity between NDCT and LDCT on different feature extractor from (a) Mayo-2016 and (b) Mayo-2020 datasets. Note that our \aename is trained on Mayo-2016 and generalizes to Mayo-2020.}
\label{fig:feature_visual}
\end{figure}

\begin{figure*}[!t]
\centering
\includegraphics[width=1\linewidth]
{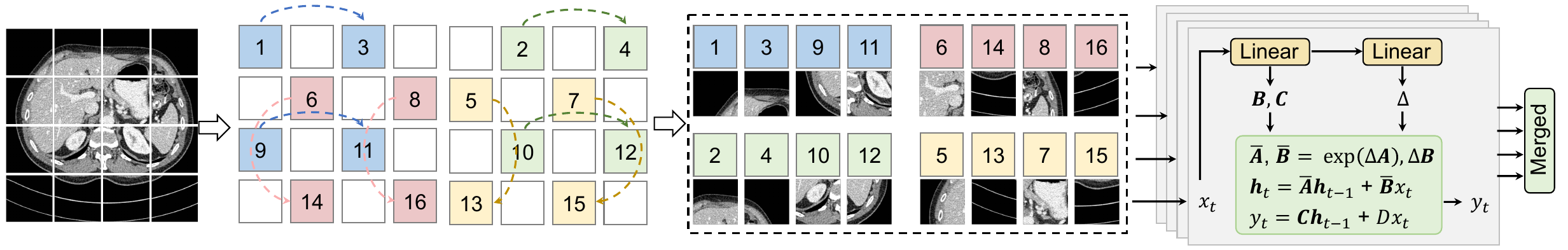}
\caption{Architecture of ES2D, which skips scan patches with a step size $2$ and partitions into selected spatial dimensional feature.}
\label{fig:es2d}
\end{figure*}

\subsection{Semantic-enhanced Efficient Denoiser (\generator)}
\label{sec:mcunet}
To effectively integrate the capabilities of VLMs, as embodied in \aename, into the denoising model, we design a 4-level U-shaped denoising backbone, shown in Fig.~\ref{fig:network}(b).
The encoder comprises the first five scales of dense features of the frozen encoder \aename and the decoder is learnable with efficient Mamba attention blocks~(EMA).
We employ this choice of encoder design to enhance NDCT-relevant local sematic, which is motivated by our finding that these dense features are remarkably noise-robust, exhibiting high cosine similarity between NDCT and LDCT representations, shown in Fig.~\ref{fig:feature_visual}.
It is a property attributable to the powerful feature extraction capabilities of VLMs.
Furthermore, Fig.~\ref{fig:feature_visual} demonstrates that the semantic information extraction capabilities of \aename, trained on one dataset, can be effectively generalized to unseen datasets. 
This significantly reduces the training and deployment costs associated with applying our model to new data.
The output features from each encoder level are added to the corresponding decoder block at the same level through skip connections.

To efficiently capture global information in the decoder, we design an EMA block that integrates an efficient SSM module~(ESSM) and a channel spatial attention (CSA) mechanism, shown in Fig.~\ref{fig:network}(b). Specifically, the input feature $\mat{F}_\mathrm{in}$ is first processed by ESSM, which extracts long-range dependencies with linear complexity. 
Simultaneously, $\mat{F}_\mathrm{in}$ is also refined through the CSA mechanism~\cite{woo2018cbam}, enhancing its global feature selection capability. 
The outputs from ESSM and CSA are then fused and passed through a feed-forward network (FFN) consisting of a convolution layer and an activation function. 
The final output of an EMA block is represented as:
\begin{equation}
\begin{aligned}
&\mat{F}^\prime_\mathrm{in} = \operatorname{ESSM}(\mat{F}_\mathrm{in})+\operatorname{CSA}(\mat{F}_\mathrm{in})+\mat{F}_\mathrm{in},\\
&\mat{F}_\mathrm{out}=\operatorname{FFN}({\mat{F}^\prime_\mathrm{in}})+\mat{F}^\prime_\mathrm{in}. 
\end{aligned}
\end{equation}

As the core design of the EMA block, the ESSM module is composed of two distinct branches, shown in Fig.~\ref{fig:network}(b). 
The input feature map, $\mat{F}_\mathrm{in}$, is first processed by a linear layer in the first branch to expand its channels. 
This is followed by a sequence of operations, including a depth-wise convolution, a SiLU activation function~\cite{silu}, an efficient 2D-selective-scan~(ES2D) module, a LayerNorm operation, which collectively refines the feature representation. 
In the second branch, the input is passed through a linear layer and subsequently activated by the SiLU function. 
The outputs of the two branches are then fused using the dot product to combine features. 
Finally, a projection operation restores the channel dimensions of the output to match those of the input, yielding an output $\mat{F}_\mathrm{out}$ with the same shape as $\mat{F}_\mathrm{in}$, which is defined as:
\begin{equation}
\begin{aligned}
&\mat{F}_1 = \operatorname{LN(ES2D(SiLU(DWConv(Linear}(\mat{F}_\mathrm{in}))))),\\
&\mat{F}_2 = \operatorname{SiLU(Linear}(\mat{F}_\mathrm{in})),\\
&\mat{F}_3 = \operatorname{Linear}(\mat{F}_1 \odot \mat{F}_2),
\end{aligned}
\end{equation}
where $ \operatorname{LN}(\cdot)$ is LayerNorm operation; and $ \operatorname{DWConv}(\cdot)$ represents the depth-wise convolution. 

As shown in Fig.~\ref{fig:es2d}, the 2D image feature is flattened into a 1D sequence with scanning along four different directions: top-left to bottom-right, bottom-right to top-left, top-right to bottom-left, and bottom-left to top-right.
Instead of scanning whole patches in the original SS2D, we skip scanning patches with a step size $s=2$ and partition them into selected spatial dimensional features, which reduces the number of processed tokens by a factor of $s^2$.
Then the long-range dependency of each sequence is captured according to the discrete state-space equation. 
Finally, all sequences are merged using summation followed by the reshape operation to recover the 2D structure.

\begin{figure*}[!t]
\centering
\includegraphics[width=1\linewidth]{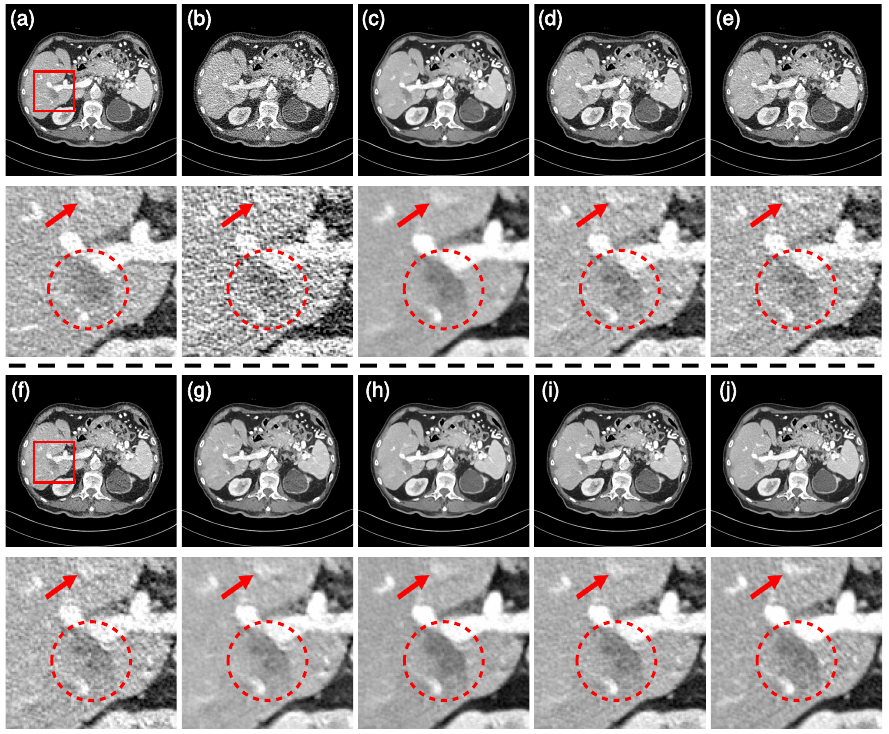}
\caption{Transverse CT images from the Mayo-2016 dataset. (a)~NDCT; (b)~LDCT; (c)~RED-CNN; (d)~EDCNN; (e)~WGAN-VGG; (f)~DU-GAN; (g)~Hformer; (h)~EASU-Net; (i)~CoreDiff; and (j)~\modelname. The ROI of the rectangle is zoomed below for better visualization. The display window is [-160, 240] HU.
}
\label{fig:test_results_spatial}
\end{figure*}

\begin{figure*}[!t]
\centering
\includegraphics[width=1\linewidth]{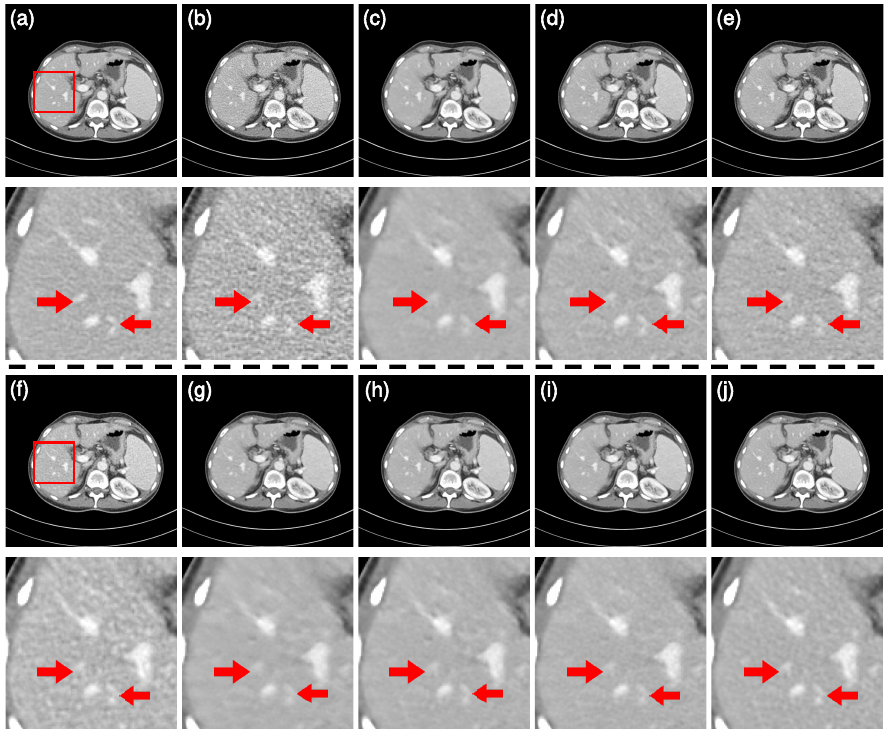}
\caption{Transverse CT images from the Mayo-2020 dataset. (a)~NDCT; (b)~LDCT; (c)~RED-CNN; (d)~EDCNN; (e)~WGAN-VGG; (f)~DU-GAN; (g)~Hformer; (h)~ESAU-Net; (i)~CoreDiff; and (j)~\modelname. The ROI of the rectangle is zoomed below for better visualization. The display window is [-160, 240] HU.}
\label{fig:test_results_spatial_2020}
\end{figure*}

\subsection{Language-engaged Dual-space Alignment (\lossname)}
\label{sec:leda}
To further enhance the supervision of semantic information, we propose the \lossname loss, which employs the frozen \aename to align the denoised CT and NDCT images in both continuous and discrete spaces, as shown in Fig.~\ref{fig:network}(c).
Specifically, following the denoising process of \generator,  the pretrained \aename encodes both the denoised CT image $\mat{\hat{y}}$ and the ground-truth NDCT image $\mat{y}$, generating the continuous features $\mat{\hat{z}}_\mathrm{e}$ and $\mat{z}_\mathrm{e}$. 
These representations are subsequently quantized into text token embeddings $\mat{\hat{z}}_\mathrm{q}$ and $\mat{z}_\mathrm{q}$, respectively.

The \lossname loss is introduced to maximize the similarities 
between denoised CT and NDCT images by minimizing the Euclidean distance in terms of both continuous perceptual features and discrete semantic token embeddings, which is defined as:
\begin{align}
    \mathcal{L}_\mathrm{\lossname}=
    \underbrace{{\|\mat{z}_\mathrm{e}}\!-\!\mat{\hat{z}}_\mathrm{e}\|_2^{2}}_{\text{continuous}}
    +\underbrace{
\|{\mat{z}_\mathrm{q}}\!-\!\mat{\hat{z}}_\mathrm{q}\|_2^{2}}_{\text{discrete}}.
\end{align}

During the training process, we combine the \lossname loss with the conventional pixel-level MSE loss to constrain the \generator. 
The final denoising loss is defined as:
\begin{align}
\mathcal{L}&=\mathcal{L}_\mathrm{MSE}+\lambda\mathcal{L}_\mathrm{\lossname},
\end{align}
where $\mathcal{L}_\mathrm{MSE}=\|{\mat{y}}\!-\!\mat{\hat{y}}\|_2^{2}$. We empirically set $\lambda$ to $0.3$.

\begin{table*}[!t]
\caption{Performance comparison (mean$\pm$STD) on the Mayo-2016 and Mayo-2020 datasets in terms of PSNR, SSIM, and FSIM. Time(s) represents the inference time per sample on the GPU.}
\label{Results_comparison}
\centering
\begin{tabular*}{1\linewidth}{@{\extracolsep{\fill}}rcccccccc}
\shline
 & \multicolumn{3}{c}{\textbf{Mayo-2016 Dataset}}          && \multicolumn{3}{c}{\textbf{Mayo-2020 Dataset}} \\
 \cline{2-4}  \cline{6-8}
\textbf{Methods} & PSNR$\uparrow$ &  SSIM$\uparrow$   &   FSIM$\uparrow$ && PSNR$\uparrow$ &  SSIM$\uparrow$   &   FSIM$\uparrow$ & Time(s)\\
\midrule

{RED-CNN~\cite{redcnn}}   & 28.69{$\pm${1.57}} & 0.8562{$\pm${0.0381}} & 0.9321{$\pm${0.0180}} && 34.03{$\pm${1.80}}  & 0.9328{$\pm${0.0217}} & 0.9686{$\pm${0.0101}} & 0.038\\   

{EDCNN~\cite{liang2020edcnn}}   & 27.90{$\pm${1.57}} & 0.8595{$\pm${0.0359}} & 0.9388{$\pm${0.0134}} && 33.62{$\pm${1.84}}  & 0.9318{$\pm${0.0230}} & 0.9707{$\pm${0.0100}} & 0.013\\   

{WGAN-VGG}~\cite{wgan-vgg}  & 27.09{$\pm${1.61}}  &  0.8558{$\pm${0.0389}}  & 0.9390{$\pm${0.0189}} && 32.82{$\pm${1.74}}  & 0.9245{$\pm${0.0252}}  & 0.9698{$\pm${0.0101}} & 0.038\\

{DU-GAN~\cite{huang2021gan}}  & 27.51{$\pm${1.58}}  & 0.8572{$\pm${0.0371}}  &  0.9407{$\pm${0.0136}}  && 32.55{$\pm${1.36}}  & 0.9203{$\pm${0.0270}}  &  0.9654{$\pm${0.0107}} & 0.038\\

{Hformer~\cite{zhang2023hformer}}   & 28.40{$\pm${1.52}} & 0.8562{$\pm${0.0373}} & 0.9308{$\pm${0.0164}}  && 33.76{$\pm${1.75}}  & 0.9311{$\pm${0.0224}} & 0.9697{$\pm${0.0099}} & 0.023\\

{ESAU-Net~\cite{chen2023ascon}}   & 28.74{$\pm${1.58}} & 0.8616{$\pm${0.0386}} & 0.9353{$\pm${0.0188}}  && 34.12{$\pm${1.93}}  & 0.9334{$\pm${0.0218}} & 0.9708{$\pm${0.0100}} & 0.035\\

{CoreDiff~\cite{corediff}}   & 28.70{$\pm${1.60}} & 0.8630{$\pm${0.0386}} &  0.9396{$\pm${0.0152}}   && 34.14{$\pm${1.80}}  & 0.9354{$\pm${0.0214}} & 0.9714{$\pm${0.0098}} & 0.255 \\

\modelname~(ours)   & \textbf{28.83}{$\pm${1.58}} & {\textbf{0.8645}}{$\pm${0.0372}} & \textbf{0.9413}{$\pm${0.0160}}  && \textbf{34.34}{$\pm${1.87}}  & \textbf{0.9361}{$\pm${0.0219}} & \textbf{0.9731}{$\pm${0.0094}} & 0.032 \\
\shline
\end{tabular*}
\end{table*}

\section{Experiments}
\label{sec:experiment}
\subsection{Datasets}
We use two publicly available datasets commonly used in LDCT denoising tasks: one released by the 2016 NIH AAPM-Mayo Clinic Low-Dose CT Grand Challenge~\cite{mccollough2017low} and a more recently released dataset~\cite{moen2021low}, which are referred to as Mayo-2016 and Mayo-2020 respectively. 
Mayo-2016 includes normal-dose abdominal CT images of 10 anonymous patients and corresponding simulated quarter-dose CT images, which has been extensively validated on clinical data~\cite{shan2019competitive}. 
Mayo-2020 provides the abdominal CT image data of 50 patients with 25\% of normal-dose.

\subsection{Implementation Details}
For the Mayo-2016 dataset, we select 4800 image pairs of $512 \times 512$ images from 8 patients for training and 1136 pairs from 2 others for testing. 
Similarly, for the Mayo-2020 dataset, we follow the practice in DU-GAN~\cite{huang2021gan} using 2,400 image pairs of $512 \times 512$ resolution from 16 patients for training, and 580 pairs from 4 remaining patients for testing.
The VLM codebook is derived from the token embedding layer of PubMedCLIP~\cite{pubmedclip}.
The encoder and decoder of VQGAN are composed of 6 ResNet blocks and 2 attention blocks~\cite{vq-gan}. 
For the pyramid semantic loss, we use a 3-layer token pyramid as shown in Fig.~\ref{fig:network}(a), with 4, 64, and 1024 tokens at each respective layer, and corresponding thresholds of 0.95, 0.9, and 0.8.
The training is performed on an NVIDIA GeForce RTX 3090, with a Hounsfield unit (HU) window range of $[-1000, 2000]$ and a batch size of 2.
To validate the generalizability of LangAE, we train it only once on the Mayo-2016 dataset and subsequently employ it in the training of denoising models for both the Mayo-2016 and Mayo 2020-datasets.
All denoising models are optimized with AdamW~\cite{loshchilov2017decoupled} ($\beta_{1}=0.9$, $\beta_{2}=0.99$, weight decay $1.0\times10^{-9}$), with the initial learning rate of $1.0\times 10^{-4}$, decreasing to $1.0\times 10^{-6}$ via cosine annealing~\cite{loshchilov2016sgdr}.
For quantitative evaluations,  we use three widely used metrics, including the peak signal-to-noise ratio (PSNR), structural similarity index (SSIM), and feature similarity index (FSIM). 
Since both datasets we used are abdominal data, we compute the quantitative metrics in the abdominal window of $[-160,240]$ HU.

\subsection{Performance Comparison}
\subsubsection{Compared Methods}
In this section, we evaluate the effectiveness of the proposed \modelname by comparing them with previous state-of-the-art (SOTA) methods, including RED-CNN~\cite{redcnn}, EDCNN~\cite{liang2020edcnn}, WGAN-VGG~\cite{wgan-vgg}, DU-GAN~\cite{huang2021gan}, Hformer~\cite{zhang2023hformer}, CoreDiff~\cite{corediff}, and ESAU-Net~\cite{chen2023ascon}. 
Among them, RED-CNN and EDCNN are two classical LDCT denoising networks that rely on CNNs. WGAN-VGG and DU-GAN are GAN-based models that employ generative adversarial training strategies to improve denoising performance. Hformer and ESAU-Net represent transformer-based approaches that incorporate self-attention mechanisms. Additionally, CoreDiff represents the latest SOTA denoising model, leveraging the diffusion model.

\subsubsection{Quantitative Evaluation}
Table~\ref{Results_comparison} presents the testing results as mean$\pm$STD over all testing slices. 
It can be observed that \modelname outperforms all previous methods, achieving the best performance across three evaluation metrics.
Specifically, as shown in Table~\ref{Results_comparison}, our method demonstrates superior performance compared to CNN-based, transformer-based, GAN-based, and diffusion-based approaches on both the Mayo-2016 and Mayo-2020 datasets.
It can be observed that \modelname outperforms the previous diffusion SOTA, CoreDiff, while requiring significantly less inference time.
These results highlight the effectiveness of the proposed \modelname and the strong generalizability of our \aename on unseen datasets.
Compared to the self-attention mechanism, which has a quadratic complexity of $\mathcal{O}(N^2)$, the ES2D in our EMA block requires only linear complexity and reduce $\mathcal{O}(N)$ in original SS2D to $\mathcal{O}(N/4)$, where $N$ represents the sequence length ($N=HW$, with $H$ and $W$ being the height and width of the feature map, respectively).

\begin{figure}[!t]
\centering
\includegraphics[width=1\linewidth]{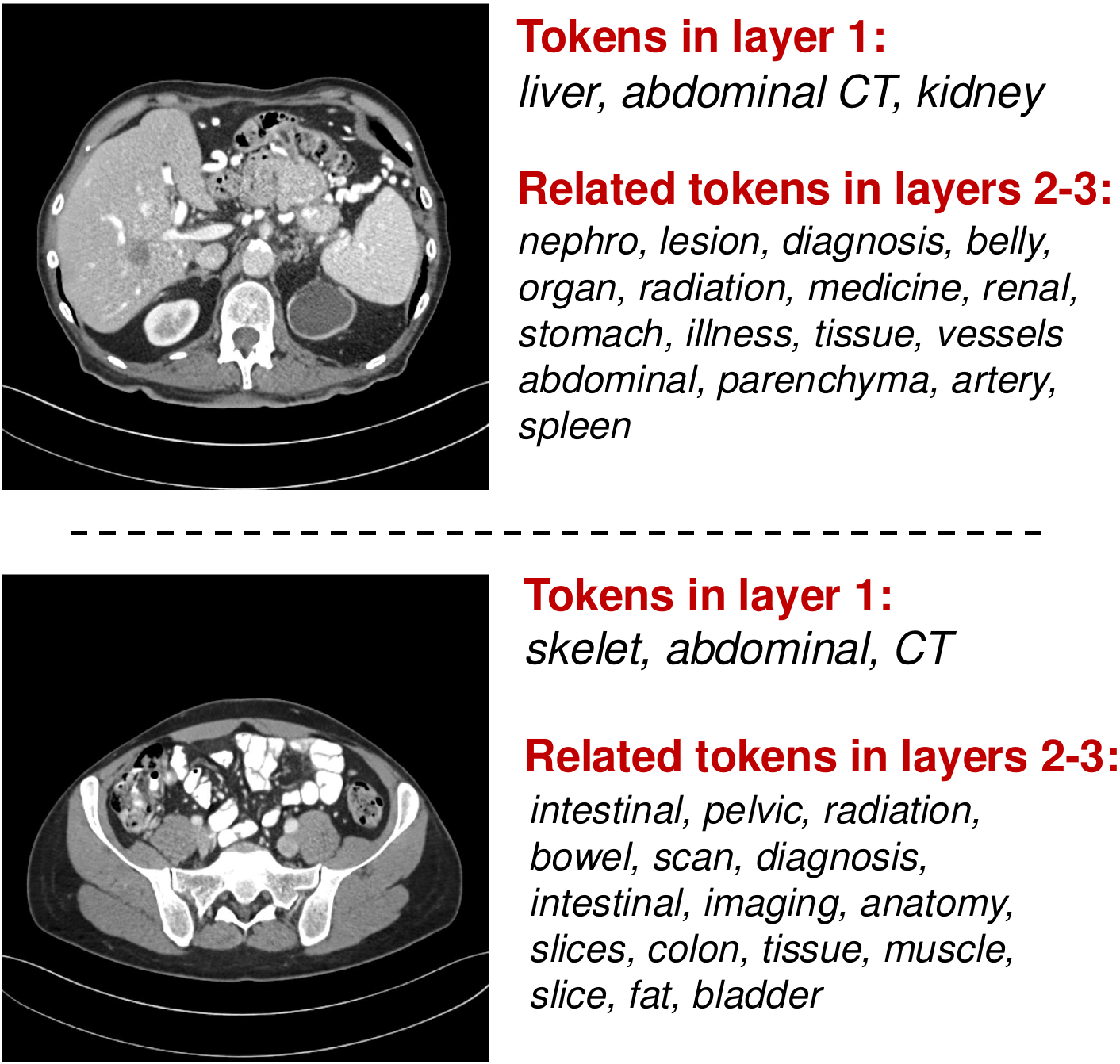}
\caption{Explainability provided by the text tokens in the 3-layer token pyramid extracted from the VLM, in which only part of quantized tokens are shown for the 2-3 layers in the quantizer.}
\label{interpretability}
\end{figure}

\subsubsection{Qualitative Evaluation}
Figs.~\ref{fig:test_results_spatial} and~\ref{fig:test_results_spatial_2020} present representative qualitative results on the Mayo-2016 and Mayo-2020 datasets, respectively.
It can be observed that, while RED-CNN effectively reduces noise, it also results in the loss of fine details. 
In contrast, Hformer and ESAU-Net maintain the shape and edge details of the lesion more effectively, due to their ability to capture long-range dependencies. 
However, these methods still suffer from over-smoothing due to the pixel-level mapping during training. 
On the other hand, GAN-based methods (\ie WGAN-VGG and DU-GAN) help mitigate this issue by better preserving texture, but they struggle with incomplete denoising and introduce velvet artifacts, likely as a result of mode collapse.
While CoreDiff improves both details and textures, it renders some blood vessels less distinct due to the stochastic nature of its diffusion process.
In contrast to the aforementioned approaches, our \modelname not only improves visual perception but also maintains clarity in critical areas like blood vessels and lesions, as shown in Figs.~\ref{fig:test_results_spatial} and \ref{fig:test_results_spatial_2020}, offering a balanced improvement in both detail preservation and visual fidelity. 
This benefits from the the semantic extraction capability of the VLM, which enhances NDCT-relevant local and global semantic in \generator and ensures semantic alignment between the denoised CT and NDCT in \lossname.
In general, the quantitative and qualitative results demonstrate the performance improvement of the proposed \modelname.

\subsection{Text Token-aided Explainability}
In addition to the performance improvement, Fig.~\ref{interpretability} shows that the \lossname loss in our \modelname can provide language-level explainability from quantized tokens.
Specifically, we visualize all tokens from the first layer of the token pyramid produced using the VLM codebook and select part of quantized tokens from the 2-3 layers.
As shown in Fig.~\ref{fig:network}(a), due to the increasing receptive field in the convolutional process, the 4 quantized tokens in the first layer capture global semantics over a larger region, while the subsequent two layers focus on more fine-grained local information.
Note that the semantic information decoded by the quantized token embeddings exhibits a certain similarity to the image content. 
For instance, the first layer aligns global concepts, such as `skelet', `CT', and `abdominal' while  2-3 layers capture more granular details, like `belly', `intestinal', and `lesion'. 
To our knowledge, this represents an initial effort to leverage the VLM for LDCT denoising. The integration of language makes the denoising model understand the content of different anatomical parts, which can improve explainability in clinical applications and make it more trustworthy for radiologists.

\begin{figure*}[!t]
\centering
\includegraphics[width=1\linewidth]{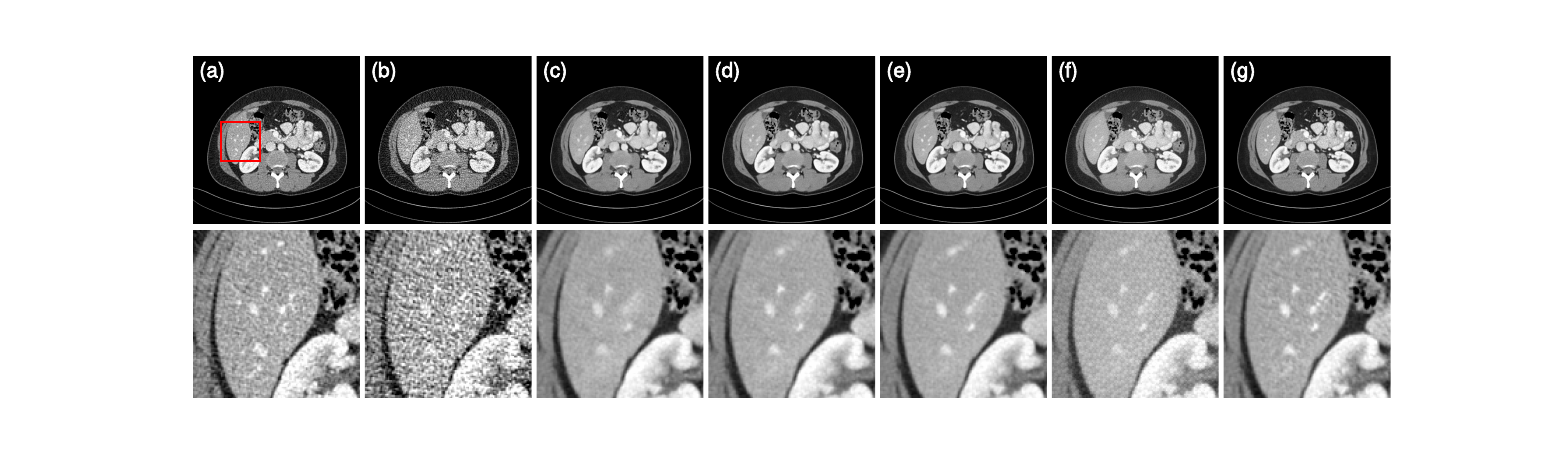}
\caption{Qualitative results of ablation studies on the EMA block and \lossname loss in \modelname. (a) NDCT; (b)~LDCT; (c)~\generator w/o EMA; (d)~\generator-R; (e)~\generator; (f) \generator+$\mathcal{L}_\mathrm{perceptual}$; and (g) \modelname.}
\label{fig:mamba_abla}
\end{figure*}

\subsection{Ablation Studies}
We conduct ablation studies on the Mayo-2016 dataset to show the effectiveness of the proposed \generator backbone and \lossname loss in our \modelname framework, respectively, as well as the effect of the VLM.

\begin{table}[!h]
\caption{Ablation results on the components of \generator.}
\label{table:mamba_abla}
\begin{tabular*}{1\linewidth}{@{\extracolsep{\fill}}lccc}
\shline
 & PSNR$\uparrow$ & SSIM$\uparrow$ & FSIM$\uparrow$\\
 \midrule
  
    \generator w/o EMA   &   28.64\std{$\pm${1.55}} & 0.8600\std{$\pm${0.0380}} & 0.9330\std{$\pm${0.0175}}
 \\
   \generator-R    &  28.70\std{$\pm${1.58}}  & 0.8604\std{$\pm${0.0385}}  &  0.9331\std{$\pm${0.0188}}  
 \\
   
  \generator   &   \textbf{28.80}\std{$\pm${1.60}} & \textbf{0.8619}\std{$\pm${0.0387}} & \textbf{0.9340}\std{$\pm${0.0192}}\\

\shline
\end{tabular*}
\end{table}

\subsubsection{Effectiveness of \generator}
To validate the effectiveness of the EMA block and \aename in \generator, we conducted experiments without using the LangDA loss and compared \generator to two designed variants, respectively: 
(1) \generator without the EMA block in the decoder (using only ResNet~\cite{he2016deep} layer), denoted as \generator without EMA, and (2) \generator with the encoder replaced by a pre-trained ResNet-18, denoted as \generator-R.
Table~\ref{table:mamba_abla} and Fig.~\ref{fig:mamba_abla} present the quantitative and qualitative results, respectively. 
It can be observed that both the EMA block and \aename effectively improve performance, enabling the generator to preserve finer details. 
This is attributed to the efficient extraction of local and global features.

\begin{table}[!h]
\caption{Ablation results on the \lossname and comparison with perceptual loss~\cite{johnson2016perceptual}.}
\label{table:loss_abla1}
\begin{tabular*}{1\linewidth}{@{\extracolsep{\fill}}lcccc}
\shline
 & PSNR$\uparrow$ & SSIM$\uparrow$ & FSIM$\uparrow$\\
 \midrule
  
  \generator   &   28.80\std{$\pm${1.60}} & 0.8619\std{$\pm${0.0387}} & 0.9340\std{$\pm${0.0192}}
 \\
   \generator+$\mathcal{L}_\mathrm{perceptual}$    &  27.30\std{$\pm${1.81}}  & 0.8581\std{$\pm${0.0393}}  &  0.9390\std{$\pm${0.0162}}  
 \\
  \modelname   & \textbf{28.83}\std{$\pm${1.58}} & {\textbf{0.8645}}\std{$\pm${0.0372}} & \textbf{0.9413}\std{$\pm${0.0160}}  
 \\

\shline
\end{tabular*}
\end{table}

\begin{table}[!h]
\caption{Ablation results on the \lossname as a plug-and-play loss.}
\label{table:loss_abla2}
\begin{tabular*}{1\linewidth}{@{\extracolsep{\fill}}lcccc}
\shline
 & PSNR$\uparrow$ & SSIM$\uparrow$ & FSIM$\uparrow$\\
 \midrule
     RED-CNN    &  28.69\std{$\pm${1.57}}  & 0.8562\std{$\pm${0.0381}}  &  0.9321\std{$\pm${0.0180}}  
 \\
 \quad+$\mathcal{L}_\mathrm{perceptual}$    &  27.47\std{$\pm${1.68}}  & 0.8542\std{$\pm${0.0385}}  &  0.9363\std{$\pm${0.0155}}  
 \\
   \quad+$\mathcal{L}_{\mathrm{\lossname}-\mathrm{C}}$    &  28.41\std{$\pm${1.57}}  & \textbf{0.8649}\std{$\pm${0.0331}}  &  \textbf{0.9421}\std{$\pm${0.0134}}  
 \\
  \quad+$\mathcal{L}_{\mathrm{\lossname}-\mathrm{D}}$   &   \textbf{28.82}\std{$\pm${1.57}}    &  0.8615\std{$\pm${0.0371}} & 0.9360\std{$\pm${0.0161}} 
 \\
  \quad+$\mathcal{L}_\mathrm{\lossname}$  &   \underline{28.73}\std{$\pm${1.46}} & \underline{0.8637}\std{$\pm${0.0357}} & \underline{0.9389}\std{$\pm${0.0173}}    \\

\shline
\end{tabular*}
\end{table}

\begin{figure*}[!t]
\centering
\includegraphics[width=1\linewidth]{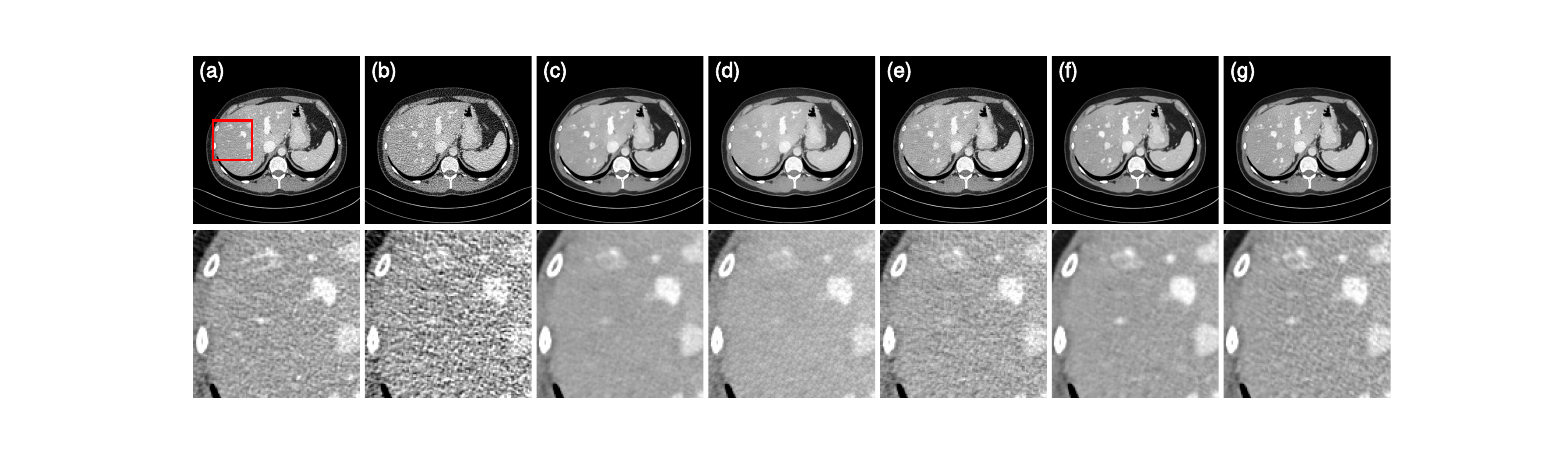}
\caption{Qualitative results on the \lossname as a plug-and-play loss in RED-CNN. (a)~NDCT; (b)~LDCT; (c)~RED-CNN; (d)~RED-CNN+$\mathcal{L}_\mathrm{perceptual}$; (e)~RED-CNN+$\mathcal{L}_\mathrm{\lossname-C}$; (f)~RED-CNN+$\mathcal{L}_\mathrm{\lossname-D}$; and (g)~RED-CNN+$\mathcal{L}_\mathrm{\lossname}$.}
\label{fig:loss_abla}
\end{figure*}

\subsubsection{Effectiveness of \lossname}
To demonstrate the effectiveness of the proposed \lossname, we compare it with the perceptual loss~\cite{wang2016perspective}, which is similar to the continuous alignment. 
The results are presented in Table~\ref{table:loss_abla1} and Fig.~\ref{fig:mamba_abla}.
As observed, the \lossname loss significantly enhances the SSIM and FSIM metrics, which prioritize texture and visual fidelity.
While \lossname yields only a slight improvement in PSNR, which rely solely on pixel-level supervision, it is widely recognized that PSNR tends to favor overly smoothed images~\cite{8839547}. 
In contrast, while the utilization of the perceptual loss enhances textural information in Fig.~\ref{fig:mamba_abla}, it compromises image feature details to some degree. 

We then validate \lossname's versatility by applying it to RED-CNN in a plug-and-play fashion.
To further explore the effectiveness of the different components in \lossname, we explore two components in our \lossname loss, where \lossname-C and \lossname-D denote the loss with only continuous space alignment and only discrete space alignment, respectively. 
Table~\ref{table:loss_abla2} and Fig.~\ref{fig:loss_abla} present the results, which show that the alignment of the continuous perceptual space retains richer texture with improved SSIM and FSIM scores, but comes at the expense of losing vessel details.  
As a supplement, \lossname-D loss, which focuses on discrete semantic space alignment, preserves finer vessel details, thereby improving the PSNR metric.
In conclusion, \lossname loss achieves a balanced improvement in both detail preservation and overall image visual fidelity.

\begin{table}[!h]
\centering
\caption{Reconstruction and denoising results constructed by different components in the \aename.}
\label{limitation}
\begin{tabular*}{1\linewidth}{@{\extracolsep{\fill}}lccc}
\shline
   & \textbf{Reconstruction quality} & \multicolumn{2}{c}{\textbf{Denoising quality}}  \\
    &    FID$\downarrow$   & PSNR$\uparrow$ & SSIM$\uparrow$ \\
 \hline
VQGAN & 58.17& 28.72 &  0.8619 \\
\quad+VLM codebook & 62.18& 28.63& 0.8591 \\
\quad\quad\quad+semantic loss & \textbf{51.79}& \textbf{28.83} & \textbf{0.8645}\\
\shline
\end{tabular*}
\end{table}

\subsubsection{Effectiveness of VLM}
The results in Table~\ref{limitation} verify the importance of autoencoder's reconstruction quality.
First, using the VLM codebook negatively affects the reconstruction results, but with semantic guidance, it performs comparably with the original VQGAN while producing semantic text tokens. 
Our semantic loss with frozen VLM codebook enhances the diversity of quantized vectors and aligns them with some diagnostic-related token embeddings (\eg~blood and tumor), so that \lossname-D makes these essential details clearer, as shown in Fig.~\ref{fig:loss_abla}. 
Second, the reconstruction quality of the \aename directly affects the performance of LDCT denoising.
We believe that improving the NDCT reconstruction quality could further elevate the performance of the \lossname loss.

\section{Discussion}
\label{sec:discussion}
\modelname achieves superior denoising performance and improved visualization results by the proposed \generator with the enhancement of local-global information and the proposed \lossname with additional semantic supervision. 
Both components benefit from the VLM introduced by \aename, strengthening the extraction of image semantic information.
We discuss the advantages of our \modelname across various denoising methods as follows.
{(\textbf{i})} Unlike the CNN and transformer-based methods, our \modelname can capture both the local and global information as well as greatly reduce the computational complexity.
{(\textbf{ii})} Unlike GAN-base models, our \modelname preserves texture details and enhances image fidelity without introducing additional artifacts caused by mode collapse. 
{(\textbf{iii})} Unlike diffusion-base models, our method does not need the multi-step reverse diffusion process, and \lossname can be directly applied to various denoising backbones, offering shorter inference times and greater versatility.
We also highlight the capability of VLM in \lossname, which guided the model to understand the denoised image with quantized text tokens, providing language-level explainability in clinical applications.

Here, we acknowledge some limitations in this work. First, the reconstruction quality of our \aename still falls short compared to the state-of-the-art models in image generation, which may result in suboptimal denoising performance.
Second, we only exhibit part of tokens from the 2-3 layers of the token pyramid because many tokens in the candidate token pools exhibit poor correlation with CT anatomical information.
Looking forward, the reconstruction quality is expected to be improved to further improve denoising performance by leveraging more powerful generation models.
Additionally, the finer alignment of the image with tokens is worthy of studying in future.
Moreover, our work highlights the potential of VLM in image restoration, paving the way for future explorations, such as employing language model directly as the denoising backbone and generating denoised images in the manner of the visual autoregressive modeling (VAR)~\cite{var}.

\section{Conclusion}
\label{sec:conclusion}
In this paper, we introduce \modelname, a novel language-driven framework for LDCT denoising that integrates VLMs to enhance supervision from NDCT. 
\aename encodes CT images into quantized tokens with anatomical information using frozen VLMs.
\generator leverages \aename to enhance NDCT local contexts and employs an EMA block for linear-complexity global feature capture.
\lossname improves NDCT supervision by enforcing perceptual and semantic consistency between denoised images and NDCT through \aename.
Extensive results have validated the effectiveness of the proposed \modelname, as well as the the generalizability of \aename. 
This is the first effort to apply VLM for LDCT denoising, which paves the way for semantic-aware medical image denoising, offering both performance gains and clinical explainability.

\section*{Acknowledgments}
All authors declare that they have no known conflicts of interest in terms of competing financial interests or personal relationships that could have an influence or are relevant to the work reported in this article.



\end{document}